\documentclass[acmsmall]{acmart}
\usepackage{booktabs,multirow,array}
\usepackage[ruled,linesnumbered]{algorithm2e}
\usepackage{bm}
\AtBeginDocument{%
  \providecommand\BibTeX{{%
    \normalfont B\kern-0.5em{\scshape i\kern-0.25em b}\kern-0.8em\TeX}}}

\setcopyright{acmcopyright}
\copyrightyear{2022}
\acmYear{2022}

\acmVolume{37}
\acmNumber{4}
\acmArticle{111}
\acmMonth{1}

\begin{document}

\title{A Unified Multi-task Learning Framework for Multi-goal Conversational Recommender Systems}

\author{Yang Deng}
\email{ydeng@se.cuhk.edu.hk}
\affiliation{%
  \institution{The Chinese University of Hong Kong}
  \city{Hong Kong SAR}
  \postcode{999077}
}

\author{Wenxuan Zhang}
\affiliation{%
 \institution{The Chinese University of Hong Kong}
 \city{Hong Kong SAR}}
 \email{wxzhang@se.cuhk.edu.hk}
 
\author{Weiwen Xu}
\affiliation{%
 \institution{The Chinese University of Hong Kong}
 \city{Hong Kong SAR}}
 \email{wwxu@se.cuhk.edu.hk}
 
\author{Wenqiang Lei}
\affiliation{%
  \institution{Sichuan University}
  \country{China}}
\email{wenqianglei@gmail.com}

\author{Tat-Seng Chua}
\affiliation{%
  \institution{National University of Singapore}
  \country{Singapore}}
\email{chuats@comp.nus.edu.sg}

\author{Wai Lam}
\affiliation{%
 \institution{The Chinese University of Hong Kong}
 \city{Hong Kong SAR}}
 \email{wlam@se.cuhk.edu.hk}

\renewcommand{\shortauthors}{Deng, et al.}

\begin{abstract}
Recent years witnessed several advances in developing multi-goal conversational recommender systems (MG-CRS) that can proactively attract users' interests and naturally lead user-engaged dialogues with multiple conversational goals and diverse topics.  
Four tasks are often involved in MG-CRS, including Goal Planning, Topic Prediction, Item Recommendation, and Response Generation. 
Most existing studies address only some of these tasks. To handle the whole problem of MG-CRS, modularized frameworks are adopted where each task is tackled independently without considering their interdependencies. 
In this work, we propose a novel Unified MultI-goal conversational recommeNDer system, namely \textbf{UniMIND}.   
In specific, we unify these four tasks with different formulations into the same sequence-to-sequence (Seq2Seq) paradigm. 
Prompt-based learning strategies are investigated to endow the unified model with the capability of multi-task learning. 
Finally, the overall learning and inference procedure consists of three stages, including multi-task learning, prompt-based tuning, and inference. 
Experimental results on two MG-CRS benchmarks (DuRecDial and TG-ReDial) show that UniMIND achieves state-of-the-art performance on all tasks with a unified model. Extensive analyses and discussions are provided for shedding some new perspectives for MG-CRS.

\end{abstract}


\maketitle

\section{Introduction}
Conversational Recommender Systems (CRS) aim to make recommendations by learning users' preferences through interactive conversations~\cite{cikm18-saur,wsdm20-ear,nips18-redial}. CRS has become one of the trending research topics for recommender systems and is gaining increasing attention, due to its natural advantage of explicitly acquiring users' real-time preferences and providing a user-engaged recommendation procedure. 
Based on different scenarios, various CRS have been proposed, either from the perspective of recommender systems, being an enhanced interactive recommender system~\cite{sigir18-crm,kdd18-q&r,wsdm20-ear}, or from the perspective of dialogue systems, being a variation of goal-oriented conversational systems~\cite{nips18-redial,emnlp19-goredial,lei2018sequicity}.

\begin{figure*}
\centering
\includegraphics[width=0.7\textwidth]{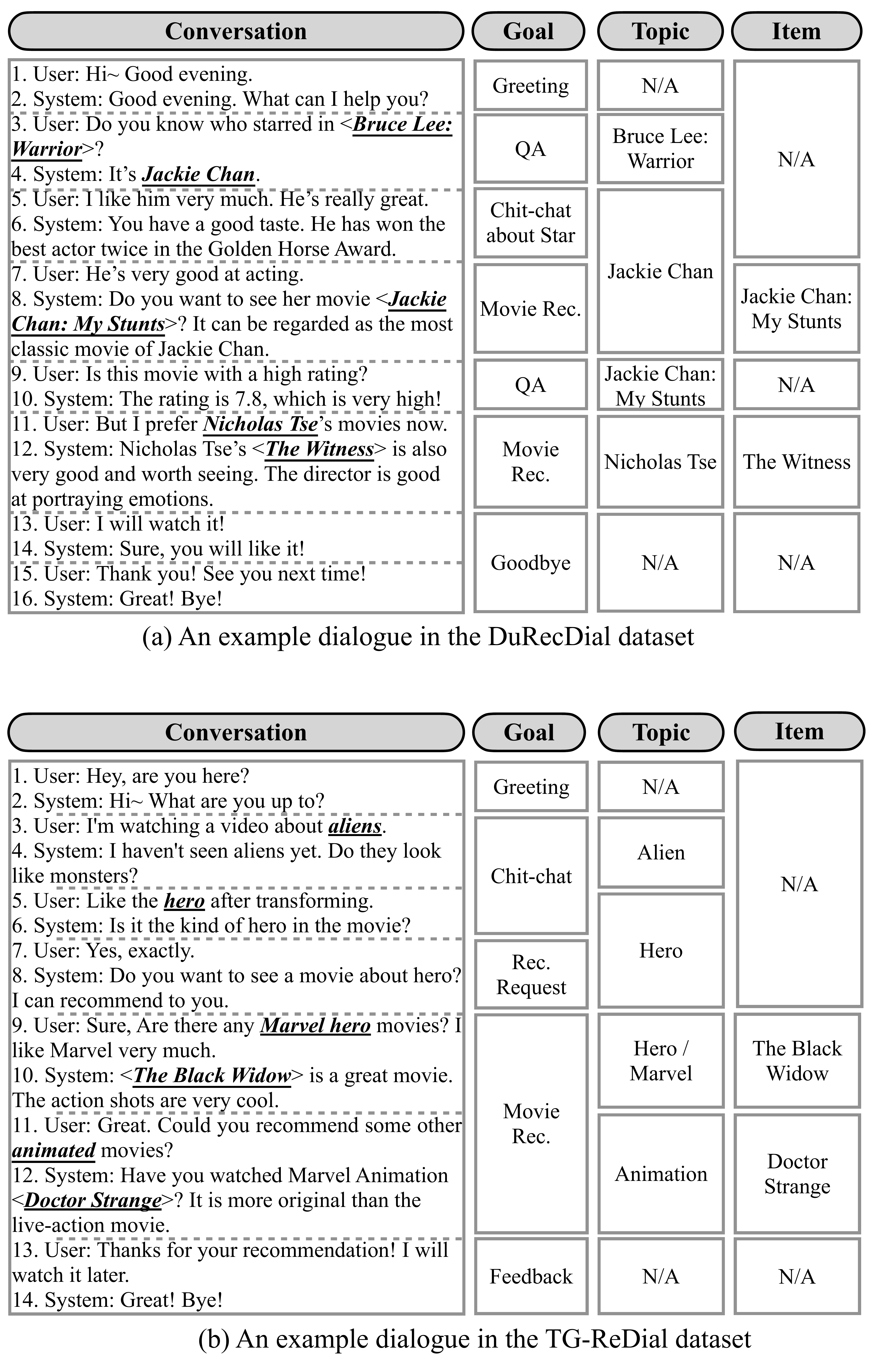}
\caption{Two illustrative examples of Multi-goal CRS from DuRecDial~\cite{durecdial} and TG-ReDial~\cite{tgredial} datasets.}
\label{example}
\end{figure*}

Most of these CRS assume that users always know what they want and the system passively and solely targets at making the successful recommendation on users' desired items. 
Recently, several efforts have been made on developing multi-goal conversational recommender systems (MG-CRS)~\cite{durecdial,tgredial} that can proactively discover users' interests and naturally lead user-engaged dialogues with multiple conversation goals or topics, not just recommendations. 
As the examples illustrated in Fig.~\ref{example}, MG-CRS is expected to dynamically adjust the conversational goals (\textit{e.g.}, from QA, Chit-chat, to Recommendation in Fig.~\ref{example}(a)) and adaptively lead the dialogue topics (\textit{e.g.}, from Alien, Hero, to Animation in Fig.~\ref{example}(b)). 
Accordingly, the problem of MG-CRS can be decomposed into  four tasks: 
\begin{itemize}
    \item \textit{Goal Planning} aims to plan the goal sequence to guide the conversation to the final recommendation.
    \item  \textit{Topic Prediction} predicts the appropriate topics\footnote{According to different applications, the topics can be specific topic classes (\textit{e.g.}, TG-ReDial), topical knowledge entities (\textit{e.g.}, DuRecDial), etc. } for the current conversational goal.
    \item \textit{Item Recommendation} provides suitable items that meet the user's need.
    \item \textit{Response Generation} produces appropriate natural language responses to users. 
\end{itemize}

Early works typically adopt modularized frameworks that address different tasks in MG-CRS with independent models. For example, \citet{durecdial} propose a CNN-based goal planning model to predict the next goal and a multi-GRU-based response model to jointly select topical knowledge and generate responses with the guidance of the predicted goal. 
\citet{tgredial} propose three separated modules based on pre-trained language models for topic prediction, item recommendation, and response generation, while the conversational goals (Chit-chat or Recommend) at each turn are pre-defined. Due to the complexity of the whole problem of MG-CRS, some recent studies tend to simplify the MG-CRS problem by either (i) assuming some information (\textit{e.g.}, the goal sequence) is priorly known~\cite{aaai21-gokc,KERS} or (ii) only performing joint learning on some of the tasks (\textit{e.g.,} topic prediction and response generation)~\cite{topicrefine}, instead of solving the whole problem of MG-CRS. 

Despite their effectiveness, there remain some limitations in the existing systems:
(i) The four tasks in MG-CRS are supposed to be closely related, while existing systems often learn each task individually without considering inter-task interactions for mutual enhancement. 
(ii) In reality, it is impossible to always have the pre-defined goal sequences or topic threads for guiding the conversation, which is also the challenge that characterizes MG-CRS from other CRS. Therefore, all tasks are expected to be learned and inferred by the system automatically.  
(iii) There are some substantial differences  among the four tasks, including various task paradigms (from multi-label classification, ranking, to text generation) and diverse required information (\textit{e.g.,} dialogue context, user profile, knowledge base, etc.). It would be time-consuming and labor-intensive to design and train several independent models for these distinct tasks.

To tackle the aforementioned challenges, we propose a novel Unified MultI-goal conversational recommeNDer system, namely \textbf{UniMIND}, which unifies four tasks in MG-CRS into the same sequence-to-sequence (Seq2Seq) paradigm and utilizes prompt-based learning strategies to endow the model with the capability of multi-task learning. 
In specific, motivated by recent successes of paradigm shifts~\cite{paradigm} on many NLP tasks, we reformulate each task in MG-CRS as a Seq2Seq problem.  Seq2Seq is a general and flexible paradigm that can handle any task whose input and output can be recast as a sequence of tokens, and better leverage the semantic relationships between input and output. 
For example, topic prediction is a multi-label classification problem, where the Seq2Seq paradigm can fully utilize the label semantics~\cite{coling18-sgm}. 
Item recommendation requires to rank a list of candidate items, where we expand the original vocabulary of the Seq2Seq model with an extra item vocabulary to capture the relationship between context words and candidate items~\cite{kbrd}. 
Furthermore, pre-trained language models (PLMs), \textit{e.g.,} GPT2~\cite{gpt2}, T5~\cite{t5}, have become the de-facto methods for dialogue generation tasks. 
In order to adapt PLMs to each task of MG-CRS, we investigate prompt-based learning strategies~\cite{prompt} to manipulate the model behavior so that the PLM itself can be used to predict the desired output and facilitate multi-task learning. 

The contributions are summarized as follows:
\begin{itemize}
    \item We propose a novel method, namely UniMIND, that tackle all tasks in MG-CRS with a unified model. To the best of our knowledge, it is the first attempt towards a unified framework for MG-CRS. 
    \item We reformulate all tasks in MG-CRS with diverse modeling paradigms into the same Seq2Seq paradigm to seamlessly unify the complex MG-CRS problem. 
    \item We investigate prompt-based learning strategies to enable the multi-task learning of all tasks in MG-CRS, and develop a special token prompting strategy that bridges the relationships within each type of information. 
    \item Experimental results on two benchmark MG-CRS datasets show that UniMIND achieves state-of-the-art performance on all tasks. Extensive analyses provide some new insights of the features in different types of dialogues, and some takeaways for future MG-CRS studies. 
\end{itemize}
\section{Related Work}
\subsection{Conversational Recommendation} 
Conversational recommender system (CRS)~\cite{crs-survey1,crs-survey2}  generally consists of two main components: a dialogue component to interact with the user and a recommender component to select items for recommendations based on user preference. According to the form of conversation, existing mainstream CRS can be divided into two groups: attribute-based CRS and open-ended CRS~\cite{rid}. 
Attribute-based CRS~\cite{cikm18-saur,sigir21-hoops,sigir21-compar,cikm21-crs,sigir21-learn2ask} asks clarification questions about the item attributes to acquire user preferences for making better recommendation. 
For these CRS, the system usually asks questions about the user’s preferences or makes recommendations multiple times, with the goal of achieving engaging and successful recommendations with fewer turns of conversations~\cite{kdd20-scpr,sigir21-crs,dasfaa21-crs}. 
Open-ended CRS~\cite{nips18-redial,emnlp19-goredial,kdd20-redial-kg,kbrd} focuses on how to understand users' preferences and intentions from their utterances and interacts with user through natural language conversations.  

All aforementioned studies on CRS typically target at a single goal, \textit{i.e.}, making successful recommendations. 
Some latest studies~\cite{tgredial,durecdial,inspired} aim at solving the problem of multi-goal conversational recommender systems (MG-CRS), which involves a sequence of goals to lead a user-engaged conversation, such as recommendation, chit-chat, QA, topic-focused dialogue, etc.  
Existing studies mainly focus on some of the tasks in MG-CRS. 
For example, \citet{aaai21-gokc} and \citet{KERS} assume that the complete goal sequence is given and study the task as a knowledge-grounded response selection problem. 
\citet{topicrefine} jointly predict the topics and generate responses without the consideration of item recommendation. 
In order to tackle the entire problem of MG-CRS, we investigate a unified framework for all the tasks.

\subsection{Pre-trained Seq2Seq Models for Dialogue}\label{sec:related_plm} 
Conventional dialogue systems~\cite{dialogue-survey} can be generally categorized into chitchat-based dialogue systems, which aim at conversing with users on open-domain topics, and task-oriented dialogue systems, which target at assisting users to accomplish certain goals. 
Recently, both kinds of dialogue systems benefit from the advances in pre-trained Seq2Seq models, \textit{e.g.}, GPT-2~\cite{gpt2} and T5~\cite{t5}. For example, DialoGPT~\cite{dialogpt} and Plato~\cite{plato} extend GPT-2~\cite{gpt2} and BERT~\cite{bert}, respectively, to pre-train on open-domain conversational data for dialogue generation. 
In task-oriented dialogue systems, several attempts have been made on leveraging pre-trained Seq2Seq models for generative dialogue state tracking~\cite{simpletod,emnlp21-dst-prompt}, or further applying multi-task dialogue pre-training over external dialogue corpora~\cite{soloist,pptod}. 
In the scope of CRS, pre-trained Seq2Seq models are typically adopted as the response generation module~\cite{tgredial,topicrefine}. 
In this work, we aim to maximize the utility of  pre-trained Seq2Seq models on all tasks in MG-CRS. 

\subsection{Paradigm Shift \& Prompt-based Learning} 
In the past years, modeling for most NLP tasks have converged to
several mainstream paradigms~\cite{paradigm}, including Classification, Matching/Ranking, Sequence Labeling, Seq2Seq, etc. 
Recent work has shown that models under some paradigms also generalize well on tasks with other paradigms. 
For example, multi-label or multi-task classification may be challenging for conventional classification modeling, where \citet{coling18-sgm} adopt the Seq2Seq paradigm to better capture interactions between the labels and \citet{icml20-class-rank} adopt the Matching paradigm to predict whether the text is matched with the label with descriptions. 
In order to better utilize powerful pre-trained language models (PLMs) with diverse pre-training paradigms, prompt-based learning~\cite{prompt} has been widely studied for shifting the target task to adaptive modeling paradigms with the PLM by using appropriate prompts. 
Text classification tasks can also be solved by PLMs with the Masked Language Modeling (MLM) paradigm~\cite{acl21-class-mlm,eacl21-class-mlm,naacl21-class-mlm} or the Seq2Seq paradigm~\cite{t5}. In addition, several attempts have been made on adapting pre-trained Seq2Seq models to the task of document ranking~\cite{emnlp20-ranking-generation,emnlp20findings-ranking-generation}. 
Some latest studies investigate paradigm shift or prompt-based learning on more diverse tasks, such as information extraction~\cite{acl22-infoext}, sentiment analysis~\cite{acl21-gabsa,emnlp21-quad}, etc.

In the field of dialogue systems, such paradigm shift techniques also bring inspiring progresses on task-oriented dialogue systems. 
As mentioned in Section~\ref{sec:related_plm}, recent state-of-the-art performance~\cite{simpletod,emnlp21-dst-prompt} on dialogue state tracking is achieved by the reformulation of such a structure prediction task as a text generation task. 
In addition, \citet{tickettalk} investigate the unified Seq2Seq paradigm for action prediction in transaction-based dialogue systems. 
However, there are more complicated subtasks with diverse paradigms in MG-CRS, varying from multi-label classification, ranking/recommendation, to text generation. 
In this work, we propose to unify all the subtasks with different paradigms in MG-CRS into Seq2Seq paradigm, and design prompt-based learning approaches for utilizing pre-trained Seq2Seq models for MG-CRS.

\section{Problem Definition}
Let $\mathcal{C}_t=\{c_1,...,c_{t-1}\}$ denote the dialogue context at the current conversation turn $t$. Correspondingly, let $\mathcal{G}_t=\{g_1,...,g_{t-1}\}$ and $\mathcal{K}_t=\{k_1,...,k_{t-1}\}$ denote the historical goal sequence and topic thread, respectively. 
The CRS maintains a pre-defined set of goals $\mathbb{G}$, topics $\mathbb{K}$ to be predicted, and a large set of items $\mathbb{V}$ to be recommended during the conversation. 
In some applications, there also exist the user profiles $\mathcal{P}_u$ for each user $u$, which can be historical item interaction data or certain personal knowledge. 
Overall, MG-CRS aims to (1) plan the next goal $g_t\in \mathbb{G}$, (2) predict the next topic $k_t\in \mathbb{K}$, (3) recommend appropriate items $v_t\in\mathbb{V}$, and (4) produce a proper response $c_t$ for the current turn $t$. 
Specifically, the problem of MG-CRS can be decomposed into the following four tasks:
\begin{itemize}
    \item \textit{\textbf{Goal Planning}}. At each turn $t$, given the dialogue context $\mathcal{C}_t$ and the goal history $\mathcal{G}_t$, MG-CRS first selects the appropriate goal $g_t\in \mathbb{G}$ to determine where the conversation goes. 
    \item \textit{\textbf{Topic Prediction}}. The second task is to predict the next conversational topics $k_t\in \mathbb{K}$ for completing the planned goal $g_t$, with respect to the dialogue context $\mathcal{C}_t$, the historical topic thread $\mathcal{K}_t$, and the user profile $\mathcal{P}_u$ (if exists). %
    \item \textit{\textbf{Item Recommendation}}. If the selected goal $g_t$ is to make recommendations, the CRS should recommend an item $v_t\in\mathbb{V}$, based on the dialogue context $\mathcal{C}_t$ and the user profile $\mathcal{P}_u$ (if exists). In general, the recommended item $v_t$ is supposed to be related to the predicted topics $k_t$. 
    \item \textit{\textbf{Response Generation}}. The end task is to generate a proper response $c_t$ concerning the predicted topics $k_t$ for completing the selected goal $g_t$. When the goal is to make recommendation, the generated response is also expected to provide persuasive reasons for the recommended item $v_t$.  
\end{itemize}

\begin{figure}
\centering
\includegraphics[width=0.95\textwidth]{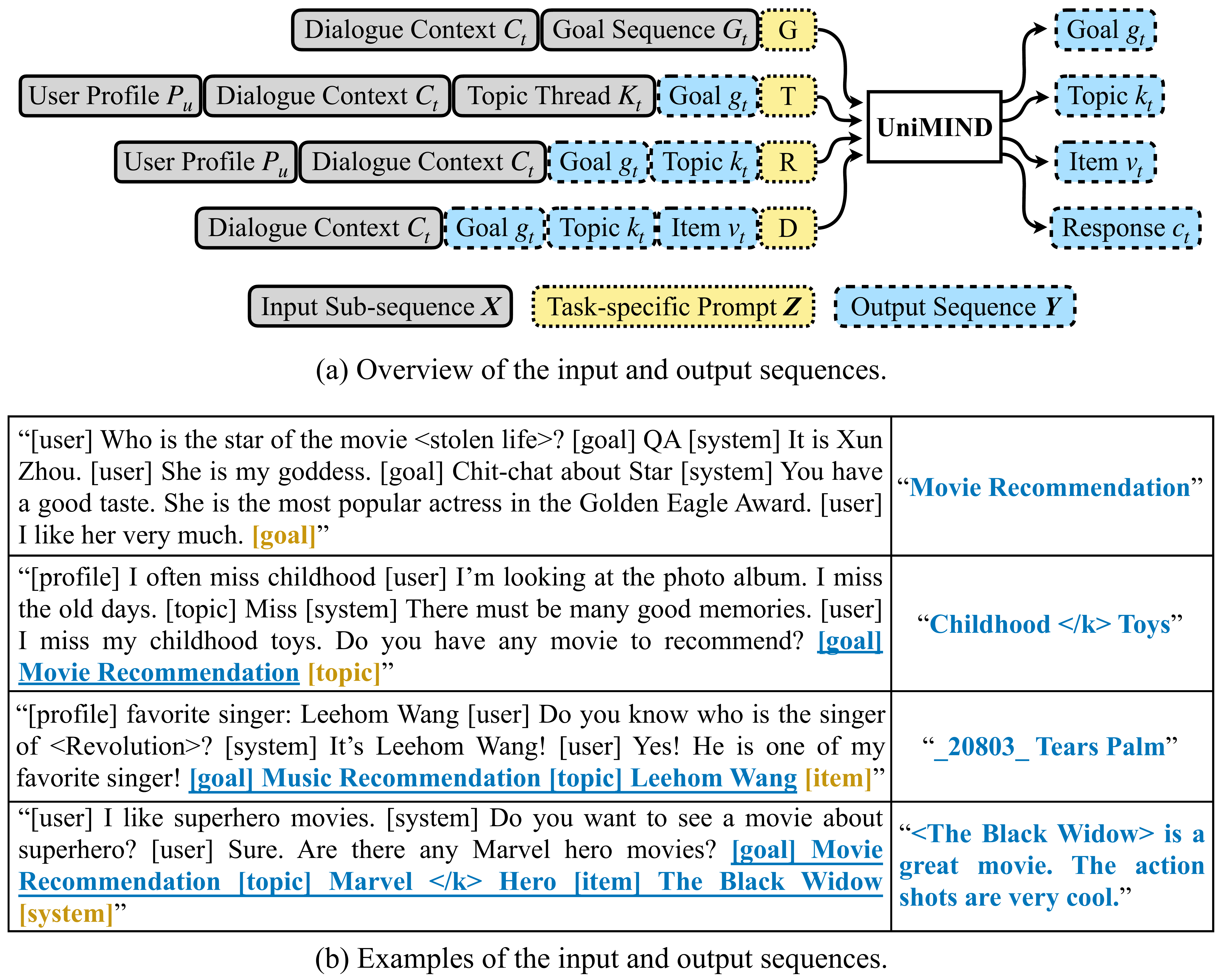}
\caption{Overview and examples of the input and output sequences for UniMIND.}
\label{overview}
\end{figure}

\section{Method}\label{sec:method}
In this section, we first describe the paradigm shift that reformulates each task in MG-CRS into the unified Seq2Seq paradigm, and then introduce the prompt-based learning strategies for the multi-task learning of all tasks. 
The overview and examples of the input and output sequences for UniMIND are illustrated in Fig.~\ref{overview}. 
Overall, the learning and inference procedure consists of three stages, including multi-task learning, prompt-based learning, and inference.

\subsection{Unified Paradigm Shifts}\label{sec:paradigm}
\subsubsection{\textbf{Goal Planning}} 
To dynamically adjust the conversation goal in the multi-goal conversational recommendation systems, \citet{durecdial} divide the task of goal planning into two sub-tasks, goal completion estimation and goal prediction. Goal completion estimation aims at estimating the probability of goal completion by performing a \textit{binary classification}. 
Goal prediction aims to predict the next goal when the previous goal is completed by performing \textit{multi-task classification}: 
\begin{equation}
    \bm{y}_\text{GP} = \mathbf{CLS}(\mathbf{Enc}(\mathcal{C}_t,\mathcal{G}_t)) \in \{0,1\}^{|\mathbb{G}|},
\end{equation}
where $\mathbf{Enc}(\cdot)$ is the encoders that encode the dialogue context and the goal history, and $\mathbf{CLS}(\cdot)$ is a multi-task classifier. $|\mathbb{G}|$ is the total number of all possible goal classes. 

We unify these two sub-tasks as one Seq2Seq task, which aims at directly generating the label of the next goal as natural language with a text generation model, \textit{i.e.,} $\mathbf{UniMIND}(\cdot)$:
\begin{gather}\label{eq:goal}
    g_t  = \mathbf{UniMIND}(\mathcal{C}_t,\mathcal{G}_t).
\end{gather}
The objective of the Goal Planning task $G$ is to maximize the sequential log-likelihood: 
\begin{gather}
    \mathcal{L}_G  = \log p(g_t|\mathcal{C}_t,\mathcal{G}_t) = \sum^{L_g}_{l=1}\log p_\theta(g_{t,l}|g_{t,<l};\mathcal{C}_t,\mathcal{G}_t),
\end{gather}
where $L_g$ denotes the target length of the generated goal label. Such a paradigm shift can alleviate the error propagation in the two-stage classification method.

\subsubsection{\textbf{Topic Prediction}}
According to different applications, the topic can be knowledge entities or specific topic classes, and there are also different corresponding solutions. 
For instance, \citet{durecdial} implicitly select the relevant knowledge triples by assigning attention weights to all candidate knowledge and then fusing them into a single vector. 
\citet{tgredial} predict the next topics by performing a \textit{text matching} task:
\begin{equation}
    {y_\text{KS}}_i = \mathbf{CLS}(\mathbf{Enc}(k_i), \mathbf{Enc}(\mathcal{C}_t, \mathcal{K}_t, \mathcal{P}_u)),
\end{equation}
where the predicted topic is ranked with the highest relevance score, \textit{i.e.}, $\arg\max_i {y_\text{KS}}_i$. 
However, these approaches may fail to handle those scenarios with a large set of topics to be predicted. Besides, the fixed number of predicted topics may also affect the cases where there is no topic needed or multiple topics are involved.  

We regard the topic prediction as an multi-label classification problem and reformulate it into the Seq2Seq paradigm, where the predicted labels are concatenated into a single sequence as the target sequence to be generated. A special token, \textit{e.g.}, ``\texttt{</k>}'', is used to separate each individual label.   
\begin{align}\label{eq:topic}
    k_t &= \mathbf{UniMIND}(\mathcal{C}_t, \mathcal{K}_t, \mathcal{P}_u, g_t),\\
    \begin{split}
        \mathcal{L}_K &= \log p(k_t|\mathcal{C}_t,\mathcal{K}_t, \mathcal{P}_u, g_t) \\
        &= \sum^{L_k}_{l=1}\log p_\theta(k_{t,l}|k_{t,<l};\mathcal{C}_t, \mathcal{K}_t, \mathcal{P}_u, g_t),
    \end{split}
\end{align}
where $L_k$ denotes the target length of generated sequence of topic labels, and $\mathcal{L}_K$ is the objective function of the Topic Prediction task $T$. The final prediction results can be recovered by splitting the sequence using ``\texttt{</k>}''. 
By doing so, the MG-CRS will be more scalable and flexible to different situations, even when the number of possible topics is extremely large and variable.  
In addition, it can be observed from Fig.~\ref{example} that the next topic is highly related to the dialogue context. And the Seq2Seq paradigm can make full use of the semantic relationships between the dialogue context and the predicted topic labels as well as among different topic labels.

\subsubsection{\textbf{Item Recommendation}}
There are two main-stream solutions for item recommendation in CRS: (i) Following traditional recommendation systems, the recommendation module~\cite{tgredial,kdd20-redial-kg} aims to rank all the items by computing the probability that recommend an item $v_i$ to the user $u$:
\begin{equation}
    {y_\text{Rec}}_i = \mathbf{CLS}(e_i,\mathbf{Enc}(\mathcal{C}_t,  \mathcal{P}_u)), 
\end{equation}
where $e_i$ is the trainable item embedding, and the recommended item is ranked with the highest recommendation probability, \textit{i.e.}, $\arg\max_i {y_\text{Rec}}_i$. 
(ii) Some studies~\cite{kbrd,rid} perform the item recommendation within an end-to-end CRS, by regarding the item set as an additional vocabulary for generation, \textit{i.e.}, each item index as a word in the expanded vocabulary. 

Motivated by the second type of approach, we extend the idea of vocabulary expansion with the candidate item set to further take into account the semantic information of the item.  
Specifically, the target sequence of the item recommendation task is composed of the item expressions in both the original vocabulary and the expanded item vocabulary. 
For example, if the target item in a sample is ``\texttt{The Witness}'' with the item index as 100, then the target sequence for this sample will be ``\texttt{\_100\_ The Witness}''. 
By doing so, not only can the model capture the relationship between context words and candidate items, but also exploit the semantic information of items.  
\begin{align}
    v_t &= \mathbf{UniMIND}(\mathcal{C}_t, \mathcal{P}_u, g_t, k_t),\\
    \begin{split}
        \mathcal{L}_R &= \log p(v_t|\mathcal{C}_t, \mathcal{P}_u, g_t, k_t) \\
        &= \sum^{L_r}_{l=1}\log p_\theta(v_{t,l}|v_{t,<l};\mathcal{C}_t, \mathcal{P}_u, g_t, k_t),
    \end{split}
\end{align}
where $L_r$ denotes the target length, and $\mathcal{L}_R$ is the objective function of the Item Recommendation task $R$. 
In the inference, we just conduct decoding for one step and rank the largest probability of the item in the expanded item vocabulary to recommend:
\begin{equation}\label{eq:item}
    v_t = \arg\max_{v^{(i)}\in\mathbb{V}} p_\theta(v^{(i)}|\texttt{[sos]};\mathcal{C}_t, \mathcal{P}_u, g_t, k_t),
\end{equation}
where \texttt{[sos]} denote the start-of-sentence token. 

\subsubsection{\textbf{Response Generation}}
Since response generation is a standard text generation problem in the Seq2Seq paradigm, it does not require any paradigm shift. 
\begin{align}\label{eq:resp}
    c_t &= \mathbf{UniMIND}(\mathcal{C}_t, g_t, k_t, v_t),\\
    \begin{split}
        \mathcal{L}_D &= \log p(c_t|\mathcal{C}_t, g_t, k_t, v_t) \\
        &= \sum^{L_d}_{l=1}\log p_\theta(c_{t,l}|c_{t,<l};\mathcal{C}_t, g_t, k_t, v_t),
    \end{split}
\end{align}
where $L_d$ denotes the target length of generated responses, and $\mathcal{L}_D$ is the objective function of the Response Generation task $D$. 

\subsection{Prompt-based Learning}
After unifying all tasks in MG-CRS into the Seq2Seq paradigm, each input and output sequence pair in each task forms an annotated data instance for training a unified encoder-decoder model. 
Inspired by prompt-based learning~\cite{prompt,t5}, we add a task-specific prompt to the original input sequence to specify which task the model should perform as well as enrich the input with task-specific information. 

As discussed in Section~\ref{sec:paradigm}, the combination of required information in the input sequence varies in different tasks. 
In order to indicate each information segment with a specific type, all sub-sequences are concatenated with special segment tokens, such as \texttt{[user]}, \texttt{[system]}, \texttt{[goal]}, \texttt{[topic]}, \texttt{[item]}, \texttt{[profile]}, etc. 
These special tokens are regarded as additional vocabulary with randomly initialized embeddings to be learned. 
For instance, the original input $X$ can be represented as:
\begin{align*}
    X_G=&\texttt{[goal]}g_1\texttt{[user]}c_1\texttt{[goal]}g_2\texttt{[system]}...\texttt{[user]}c_{t-1};\\
    X_T=&\texttt{[profile]}\mathcal{P}_u\texttt{[topic]}k_1\texttt{[user]}c_1\texttt{[topic]}k_2\texttt{[system]}...\texttt{[user]}c_{t-1}\texttt{[goal]}g_t;\\
    X_R=&\texttt{[profile]}\mathcal{P}_u\texttt{[user]}c_1\texttt{[system]}...\texttt{[user]}c_{t-1}\texttt{[goal]}g_t\texttt{[topic]}k_t;\\
    X_D=&\texttt{[user]}c_1\texttt{[system]}...\texttt{[user]}c_{t-1}\texttt{[goal]}g_t\texttt{[topic]}k_t\texttt{[item]}v_t.
\end{align*}
Specific examples are presented in Fig.~\ref{overview}(b).  
The input for prompt-based learning is composed of the original input sequence $X$ and a task-specific prompt $Z$.  We investigate two types of task-specific prompts, namely \textit{Natural Language Prompt} and \textit{Special Token Prompt}. 

\textbf{Natural Language Prompt (UniMIND$_\text{N}$).} Similar to T5~\cite{t5}, the natural language prompt employs a guidance sentence to indicate each task as follows: 
\begin{align*}
Z_G&=``\texttt{Plan the next goal:}";\\ Z_T&=``\texttt{Predict the next topic:}";\\
Z_R&=``\texttt{Recommend an item:}";\\
Z_D&=``\texttt{Generate the response:}". 
\end{align*}

\textbf{Special Token Prompt (UniMIND$_\text{S}$).} Another realization of the task-specific prompt is based on the special segment tokens as follows:
\begin{align*}
Z_G&=\texttt{[goal]};\quad Z_T=\texttt{[topic]};\\ Z_R&=\texttt{[item]};\quad Z_D=\texttt{[system]},
\end{align*}
which can be categorized into the family of \textit{continuous prompts}~\cite{prompt}. 
Originally, the special segment tokens are designed to indicate the beginning of each type of information in the input sequence. Therefore, the learned representations are supposed to preserve the knowledge of the different forms or patterns of each type of information. For example, \texttt{[goal]} and \texttt{[topic]} can separate two groups of classification labels, while \texttt{[system]} can also distinguish the speaker features of the system responses from the user utterances. 

\begin{algorithm}[t]
\caption{Learning and Inference Procedure}
\label{algo}
\KwIn{Train-set $\mathcal{D}_1=\{\mathcal{D}_G,\mathcal{D}_T,\mathcal{D}_R,\mathcal{D}_D\}=\{(X,Y,Z)_i\}_{i=1}^{|\mathcal{D}_1|}$; 
Test-set $\mathcal{D}_2=\{(\mathcal{C},\mathcal{G},\mathcal{K},\mathcal{P}_u)_i\}_{i=1}^{|\mathcal{D}_2|}$; Pre-trained parameters $\theta$; Training epoch number $e_1$; Fine-tuning epoch number $e_2$; Loss function $\mathcal{L}_\theta$;} 
// Multi-task Training\;
\For{$\mathit{epoch} = 1, 2, \ldots , e_1$}{
\For{Batch $B$ in Shuffle($\mathcal{D}_1$)}{
	Use $B=\{(X,Y,Z)_i\}_{i=1}^{|B|}$ to optimize $\mathcal{L}_\theta$ in Eq.~(\ref{eq:loss})\;
}
} 
// Prompt-based Learning\;
\For{Task $t$ in $[G,T,R,D]$}{
    $\theta_t = \theta$\;
    \For{$\mathit{epoch} = 1, 2, \ldots , e_2$}{
        \For{Batch $B$ in Shuffle($\mathcal{D}_t$)}{
            Use $B=\{(X,Y,Z)_i\}_{i=1}^{|B|}$ to optimize $\mathcal{L}_{\theta_t}$ in Eq.~(\ref{eq:loss})\;
        }
    }
}
// Inference\;
\For{Sample $d_i$ in $\mathcal{D}_2$}{
    $d_i = (\mathcal{C},\mathcal{G},\mathcal{K},\mathcal{P}_u)_i$\;
    Obtain $g_i$, $k_i$, $v_i$, $c_i$ in order via Eq.~(\ref{eq:goal}, \ref{eq:topic}, \ref{eq:item}, \ref{eq:resp}) using tuned model $\theta_G$, $\theta_T$, $\theta_R$, $\theta_D$, respectively\;
}
\KwOut{Models $\theta_G$, $\theta_T$, $\theta_R$, $\theta_D$; Prediction results $\{g,k,v,c\}_i^{|\mathcal{D}_2|}$.}

\end{algorithm}

\subsection{Multi-task Training and Inference}
The proposed UniMIND model is initialized with weights from a pre-trained LM in an encoder-decoder fashion, \textit{e.g.}, BART~\cite{bart} or T5~\cite{t5}. 
The overall learning and inference procedure of UniMIND is presented in Algorithm~\ref{algo}, which consists of three stages, including multi-task training, prompt-based learning, and inference. 

In the multi-task training stage, the model is trained to perform all tasks in MG-CRS with all training data. 
Given the training sample $(X,Y,Z)$, the objective $\mathcal{L}_\theta$ is to maximize the log-likelihood: 
\begin{equation}\label{eq:loss}
    \mathcal{L}_\theta = \sum_{l=1}^L \log p_\theta(y_l|y_{<l};X,Z), 
\end{equation}
where $L$ denotes the length of the target sequence. When applying the trained UniMIND to a specific task, we first use the same objective function as in the multi-task training stage, \textit{i.e.}, Eq.~(\ref{eq:loss}), to perform task-specific prompt-based learning. 
In the inference stage, given a data sample $d = (\mathcal{C},\mathcal{G},\mathcal{K},\mathcal{P}_u)$, we perform the four tasks in order using corresponding tuned models. 
Finally, the predicted output contains the next goal $g$, the next topic $k$, the recommended item $v$, and the generated response $c$.  
\section{Experimental Setups}\label{sec:exp_setup}

\subsection{Research Questions}
The empirical analysis targets the following research questions: 
\begin{itemize}
    \item \textbf{RQ1}: How is the performance of the proposed method on the end task of MG-CRS, \textit{i.e.}, Response Generation, compared to existing methods? 
    \item \textbf{RQ2}: How is the performance of the proposed method on each sub-task of MG-CRS, including Goal Planning, Topic Prediction, and Item Recommendation, compared to existing methods? 
    \item \textbf{RQ3}: How does the three-stage training and inference procedure as well as each sub-task  contribute to the overall performance? 
    \item \textbf{RQ4}: What is the difference between multi-goal conversational recommender systems and traditional conversational recommender systems?
\end{itemize}

\begin{table}
    \caption{Satistics of datasets.}
    \centering
    \begin{tabular}{lrr}
    \toprule
        Dataset & TG-ReDial & DuRecDial \\
        \midrule
        \#Dialogues&10,000&10,190\\
        train/dev/test&8,495/757/748&6,618/946/2,626\\
        \#Utterances&129,392&155,477\\
        \#Goals&8&21\\
        \#Topics/Entities&2,571&701\\
        \#items&33,834&11,162\\
    \bottomrule
    \end{tabular}
    \label{dataset}
\end{table}

\subsection{Datasets}
We conduct the experiments on two multi-goal conversational recommendation datasets, namely DuRecDial\footnote{\url{https://github.com/PaddlePaddle/Research/tree/48408392e152ffb2a09ce0a52334453e9c08b082/NLP/ACL2020-DuRecDial}}~\cite{durecdial} and TG-ReDial\footnote{\url{https://github.com/RUCAIBox/TG-ReDial}}~\cite{tgredial}. The dataset statistics are presented in Table~\ref{dataset}. We adopt the same train/dev/test split in the original datasets. 
\begin{itemize}
    \item \textbf{DuRecDial} is a goal-oriented knowledge-driven conversational recommendation dataset, which contains dialogues across multiple domains, including movie, music, restaurant, etc. The conversational goals and knowledge entities at each conversation turn are given. There are 21 types of conversational goals. We treat the knowledge entities as conversational topics, as the example in Fig.~\ref{example}(a). The dataset provides the user profiles with historical item interactions and the knowledge base related to each dialogue. 
    \item \textbf{TG-ReDial} is a topic-guided conversational recommendation dataset in the movie domain. We regard the labeled actions in the TG-ReDial dataset as the conversational goals at each turn. There are 8 types of conversational goals in TG-ReDial. An example is shown in Fig.~\ref{example}(b). The user profiles with historical item interactions are also given. 
\end{itemize}

\subsection{Baselines \& Evaluation Metrics}

\subsubsection{Response Generation} 
Since the original settings of TG-ReDial and DuRecDial have some differences, we compare to different groups of response generation baselines. For TG-ReDial, we consider the following baselines for comparisons: 
\begin{itemize}
    \item \textbf{ReDial}~\cite{nips18-redial} and \textbf{KBRD}~\cite{kbrd} apply the hierarchical RNN and the Transformer architecture, respectively, for response generation in conversational recommendation. 
    \item \textbf{Trans.}~\cite{transformer} and \textbf{GPT-2}~\cite{gpt2} are two general text generation baselines, which are also adopted for the evaluation on the DuRecDial dateset. 
    \item \textbf{Union}~\cite{tgredial} and \textbf{TopRef.}~\cite{topicrefine} employ GPT-2 to generate the response conditioned on the predicted topic or the recommended item. 
\end{itemize}
For DuRecDial, we compare to the following baselines: 
\begin{itemize}
    \item \textbf{seq2seq}~\cite{seq2seq} and \textbf{PGN}~\cite{pgn} are two classic text generation methods. For seq2seq and Transformer, we also report the knowledge-grounded performance, \textit{i.e.}, seq2seq+kg and Trans.+kg. 
    \item \textbf{PostKS}~\cite{postks} is a knowledge-grounded response generation method.
    \item \textbf{MGCG}\footnote{Following previous studies~\cite{aaai21-gokc,KERS}, we only compare our model with MGCG\_G, since MGCG\_R is a retrieval-based dialogue model.}~\cite{durecdial} adopts multi-type GRUs to encode the dialogue context, the goal sequence, and the topical knowledge and uses another GRU to generate responses.
    \item \textbf{GOKC}~\cite{aaai21-gokc} and \textbf{KERS}~\cite{KERS} both assume that the goal sequence is given and focus on the knowledge-grounded response generation problem.  Therefore, we also report their performance without the given goal sequence (``- w/o goal''). 
\end{itemize}
For both datasets, we also adopt the vanilla \textbf{BART}~\cite{bart} as a baseline.
Following previous studies~\cite{durecdial,tgredial,aaai21-gokc,KERS}, we adopt word-level F1, BLEU, Distinct scores (Dist), and Perplexity (PPL) as automatic evaluation metrics. 

\subsubsection{Goal Planning} 
We compare to three text classification baselines: 
\begin{itemize}
    \item \textbf{MGCG}~\cite{durecdial} employs two CNN-based classifiers to perform two sub-tasks, goal completion estimation and goal prediction. 
    \item \textbf{BERT} uses the dialogue context as the input for performing a  classification task to predict the next goal.
    \item \textbf{BERT+CNN} combines the obtained representations from MGCG and BERT to predict the next goal. 
\end{itemize}
We adopt Macro-averaged Precision (P), Recall (R), and F1 as the evaluation metrics. 

\subsubsection{Topic Prediction} 
We compare to five methods: 
\begin{itemize}
    \item \textbf{MGCG}~\cite{durecdial} adopts multi-type GRUs to encode the dialogue context, the historical topic sequence, and the user profile for performing a text matching task to rank the candidate topics.
    \item \textbf{Conv/Topic/Profile/-BERT}~\cite{tgredial} utilizes BERT to encode historical utterances/topics/user profiles to rank the candidate topics, respectively.
    \item \textbf{Union}~\cite{tgredial} combines the obtained representations from Conv/Topic/Profile-BERT. 
\end{itemize}
Following \citet{tgredial}, we adopt Hit@$k$ as the evaluation metrics for ranking all the possible topics. Besides, we also report Micro-average P, R, and F1 of all the test instances. Since there are some responses that have no topic, if the prediction on a response with no gold labels is also empty, it means the model performs well and we set the P, R, F1 to 1, otherwise 0. 

\subsubsection{Item Recommendation} 
We compare to five recommendation baselines: 
\begin{itemize}
    \item \textbf{GRU4Rec}~\cite{gru4rec} and \textbf{SASRec}~\cite{sasrec} apply GRU and Transformer, respectively, to encode user interaction history without using conversation data.
    \item \textbf{TextCNN}~\cite{textcnn} and \textbf{BERT}~\cite{bert} adopt a CNN-based model and a BERT-based model, respectively, to encode the dialogue context without using historical user interaction data.
    \item \textbf{Union}~\cite{tgredial} combines the learned representations from SASRec and BERT to make recommendations. 
\end{itemize}
Following \citet{tgredial}, we adopt NDCG@$k$ and MRR@$k$ as the evaluation metrics.

\subsection{Implementation Details}
Most results of the baselines are reported in previous works. For reproducing some additional results, we implement those baselines with the open-source CRS toolkit, CRSLab\footnote{\url{https://github.com/RUCAIBox/CRSLab}}~\cite{crslab}. In order to make a fair comparison with other baselines, we choose BART$_\text{base}$ as the pre-trained Seq2Seq model for UniMIND, which shares a similar number of model parameters with BERT$_\text{base}$ and GPT-2.  
The pre-trained weights of BART$_\text{base}$ are initialized using the Chinese BART\footnote{\url{https://huggingface.co/fnlp/bart-base-chinese}}~\cite{bart-chinese}. 
We use the same hyper-parameter settings for these two datasets. The learning rate and the weight decay rate are set to be 5e-5 and 0.01, respectively. The max source sequence length and the max target sequence length are 512 and 100, respectively. 
For DuRecDial dataset, we extract the knowledge triples concerning the predicted topical entities from the given knowledge base, and regard these knowledge triples as the complete topic context for the response generation task, whose maximum length is set to be 256. 
We train all the baselines up to 20 epochs. For the proposed method, UniMIND, we conduct multi-task training for 15 epochs and prompt-based learning for 5 epochs.\footnote{The code will be publicly released  via \url{https://github.com/dengyang17/UniMIND}.}

\section{Experimental Results}\label{sec:exp}
We evaluate the proposed method on both the end-to-end response generation and each sub-task of MG-CRS. 

\subsection{Evaluation on Response Generation (RQ1)}
We conduct both automatic and human evaluation on the response generation task. 

\subsubsection{\textbf{Automatic Evaluation}}

\begin{table}
    \caption{End-to-end Evaluation of Response Generation on TG-ReDial. The mark $\bigcirc$ denotes that the ground-truth labels are given on both training and testing, $\checkmark$ denotes that the ground-truth labels are given for training only, and $\times$ for none. $^\dagger$ indicates statistically significant improvement ($p$<0.05) over \underline{the best baseline}.}
    \centering
    \begin{tabular}{lcccccc}
    \toprule
        \multirow{2}{*}{Model}&\multicolumn{2}{c}{Ground-truth}&\multicolumn{4}{c}{Evaluation Metrics}\\ 
        \cmidrule(lr){2-3}\cmidrule(lr){4-7}
        &Goal&Topic& F1 & BLEU-1/2 & Dist-2 & PPL \\
        \midrule
        ReDial$^*$~\cite{nips18-redial}&$\times$&$\times$&-&0.177/0.028&0.025&81.61\\
        KBRD$^*$~\cite{kbrd}&$\times$&$\times$&-&0.221/0.028&0.025&28.02\\
        Trans.$^*$~\cite{transformer}&$\times$&$\times$&-&0.287/0.071&0.083&32.86\\
        GPT-2$^*$~\cite{gpt2}&$\times$&$\times$&-&0.279/0.066&0.094&13.38\\
        Union~\cite{tgredial}&$\bigcirc$&$\checkmark$&-&0.280/0.065&0.094&\underline{7.22}\\
        TopicRef.~\cite{topicrefine}&$\bigcirc$&$\checkmark$&-&\underline{0.294/0.086}&-&-\\
        BART~\cite{bart}&$\times$&$\times$&\underline{32.80}&0.291/0.070&\underline{0.097}&7.59\\
        \midrule
        \textbf{UniMIND}$_\text{N}$&$\checkmark$&$\checkmark$&35.40$^\dagger$&0.310/0.089$^\dagger$&\textbf{0.200}$^\dagger$&6.81$^\dagger$\\
        \textbf{UniMIND}$_\text{S}$&$\checkmark$&$\checkmark$&\textbf{35.62}$^\dagger$&\textbf{0.314/0.090}$^\dagger$&0.198$^\dagger$&\textbf{5.22}$^\dagger$\\
    \bottomrule
    \multicolumn{5}{l}{$^*$ Results reported from \citet{tgredial}.} 
    \end{tabular}
    \label{exp:resp_tg}
\end{table}

\begin{table}
    \caption{End-to-end Evaluation of Response Generation on DuRecDial. The mark $\bigcirc$ denotes that the ground-truth labels are given on both training and testing, $\checkmark$ denotes that the ground-truth labels are given for training only, and $\times$ for none. $^\dagger$ indicates statistically significant improvement ($p$<0.05) over \underline{the best baseline}.}
    \centering
    \begin{tabular}{lcccccc}
    \toprule
        \multirow{2}{*}{Model}&\multicolumn{2}{c}{Ground-truth}&\multicolumn{4}{c}{Evaluation Metrics}\\ 
        \cmidrule(lr){2-3}\cmidrule(lr){4-7}
        &Goal&Topic& F1 & BLEU-1/2 & Dist-2 & PPL \\
        \midrule
        seq2seq$^*$~\cite{seq2seq}&$\times$&$\times$&26.08&0.188/0.102&0.013&22.82\\
        seq2seq+kg$^{**}$~\cite{seq2seq}&$\times$&$\checkmark$&24.52&0.165/0.079&0.013&24.75\\
        PGN$^*$~\cite{pgn}&$\times$&$\times$&33.95 &0.243/0.161& 0.039& 24.28\\
        PostKS$^*$~\cite{postks}&$\times$&$\checkmark$&39.87& 0.343/0.244& 0.056& 15.32\\
        MGCG~\cite{durecdial}&$\checkmark$&$\checkmark$&42.04& 0.362/0.252& 0.081 &14.89\\
        Trans.$^{**}$~\cite{transformer}&$\times$&$\times$&41.79& 0.393/0.288&0.050&9.78\\
        Trans.+kg$^{**}$~\cite{transformer}&$\times$&$\checkmark$&44.73& 0.419/0.318&0.055&9.40\\
        GPT-2~\cite{gpt2}&$\times$&$\times$&47.01& 0.392/0.295& 0.165& 15.56\\
        GOKC~\cite{aaai21-gokc}&$\bigcirc$&$\checkmark$&47.28& 0.413/0.318& \underline{0.084}& 11.38\\
        - w/o goal&$\times$&$\checkmark$&45.59&0.401/0.303& 0.081& 12.45\\
        KERS~\cite{KERS}&$\bigcirc$&$\checkmark$&\underline{50.47}&\underline{0.463/0.362}&0.079&\underline{8.34}\\
        - w/o goal&$\times$&$\checkmark$&48.95&0.450/0.351& 0.082& 8.76\\
        BART~\cite{bart}&$\times$&$\times$&48.41&0.418/0.328&0.049&8.72\\
        \midrule
        \textbf{UniMIND}$_\text{N}$&$\checkmark$&$\checkmark$&\textbf{52.19}$^\dagger$&\textbf{0.479/0.398}$^\dagger$&0.079&\textbf{6.63}$^\dagger$\\
        \textbf{UniMIND}$_\text{S}$&$\checkmark$&$\checkmark$&51.87$^\dagger$&0.477/0.397$^\dagger$&\textbf{0.086}&6.69$^\dagger$\\
    \bottomrule
    \multicolumn{7}{l}{$^*$ Results reported from \citet{aaai21-gokc}. $^{**}$ Results reported from \citet{KERS}.}
    \end{tabular}
    \label{exp:resp_du}
\end{table}

Table~\ref{exp:resp_tg} and Table~\ref{exp:resp_du} summarize the experimental results on the end task of MG-CRS, \textit{i.e.}, Response Generation,  with different conversational recommender systems. 
Most of the baseline systems simplify the whole MG-CRS problem by assuming the conversational goals are pre-defined ($\bigcirc$) at each turn or ignoring some tasks ($\times$), including the current state-of-the-art methods on both datasets, \textit{i.e.,} TopicRef. and KERS. 
Given pre-defined goals, the systems perform much better than their original counterparts, \textit{e.g.}, GOKC and KERS, indicating the importance of the conversational goals in MG-CRS. 
Besides, simply adapting pre-trained language models to response generation in MG-CRS fails to achieve a promising performance improvement, \textit{e.g.}, GPT-2 and BART. 
Finally, UniMIND not only achieves the state-of-the-art performance on the content preservation metrics (F1, BLEU) but also has a promising performance on diversity (Dist) and fluency (PPL) on both datasets. 

Overall, the experimental results provide the answer to \textbf{RQ1}: \textit{UniMIND substantially and consistently outperforms existing strong baselines, including those baselines with pre-defined goals, with a noticeable margin on Response Generation, which is the end task of MG-CRS.}

\subsubsection{\textbf{Human Evaluation}}

\begin{table}
    \caption{Human Evaluation of Response Generation.}
    \centering
    \begin{tabular}{lcccc}
    \toprule
        Model & Fluency & Informativeness & Appropriateness & Proactivity \\
        \midrule
        GPT-2&1.36&1.39&1.25&1.71\\
        Union&1.31&1.24&1.58&1.80\\
        BART&1.81&1.31&1.40&1.77\\
        \textbf{UniMind}$_\text{N}$&1.93&1.52&1.70&\textbf{1.96}\\
        \textbf{UniMind}$_\text{S}$&\textbf{1.94}&\textbf{1.62}&\textbf{1.72}&\textbf{1.96}\\
        \midrule
        Human&1.98&1.90&1.99&1.98\\
    \bottomrule
    \end{tabular}
    \label{exp:human}
\end{table}

We conduct human evaluation to evaluate the generated response from four aspects: 
\begin{itemize}
    \item \textbf{Fluency}: how fluent and coherent the generated response is? 
    \item \textbf{Informativeness}: how rich is the generated response in information?  
    \item \textbf{Appropriateness}: is the generated response appropriate for the current topic?  
    \item \textbf{Proactivity}: how well does the generated response proactively complete the current goal? 
\end{itemize}

We randomly sample 100 dialogues from TG-ReDial and compare their responses produced by four methods (GPT-2, Union, BART, and UniMIND)\footnote{The generated responses of GPT-2 and Union on the TG-ReDial dataset are provided by \url{https://github.com/RUCAIBox/TG_CRS_Code}.}. 
Three annotators are asked to score each generated response with \{0: bad, 1: ok, 2: good\}. These annotators are all well-educated research assistants. As the results presented in Table~\ref{exp:human}, UniMIND consistently outperforms these strong baselines from different aspects of human evaluation. 
It is noteworthy that the scores on \textit{Appropriateness} and  \textit{Proactivity} are substantially improved by UniMIND, which demonstrates that UniMIND can effectively lead a proactive conversation with appropriate content. 
However, compared with the reference responses (Human), there is still much room for improvement on \textit{Informativeness} and \textit{Appropriateness} of the responses generated by UniMIND.  

The human judgements further support the above answer to the \textbf{RQ1}: \textit{The responses generated by the proposed method preserve a higher degree of fluency, informativeness as well as explicitly reflect the target conversational topics and lead a proactive conversation, which contributes to a higher overall quality. }

\subsection{Evaluation on Each Task (RQ2)}
\subsubsection{\textbf{Evaluation on Goal Planning}}

\begin{table}
    \caption{Evaluation of Goal Planning. $^\dagger$ indicates statistically significant improvement ($p$<0.05) over \underline{the best baseline}.}
    \centering
    \begin{tabular}{lcccccc}
    \toprule
        \multirow{2}{*}{Model} & \multicolumn{3}{c}{TG-ReDial} & \multicolumn{3}{c}{DuRecDial} \\
        \cmidrule(lr){2-4} \cmidrule(lr){5-7}
        & P & R & F1 & P & R & F1\\
        \midrule
        MGCG&0.7546&0.8093&  0.7794&0.5739& 0.6324& 0.5787\\
        BERT&0.8742&0.9004&  0.8858&0.9174& 0.9337& 0.9187\\
        BERT+CNN&\underline{0.8777}& \underline{0.9182}& \underline{0.8971}& \underline{0.9248}& \underline{0.9357}& \underline{0.9229}\\
        \midrule
        \textbf{UniMIND}$_\text{N}$&0.8879$^\dagger$& 0.9403$^\dagger$& 0.9122$^\dagger$&\textbf{0.9327}$^\dagger$& \textbf{0.9466}$^\dagger$& \textbf{0.9357}$^\dagger$\\
        \textbf{UniMIND}$_\text{S}$&\textbf{0.8887}$^\dagger$& \textbf{0.9425}$^\dagger$& \textbf{0.9137}$^\dagger$&0.9326$^\dagger$& 0.9369$^\dagger$& 0.9335$^\dagger$\\
    \bottomrule
    \end{tabular}
    \label{exp:goal}
\end{table}

Table~\ref{exp:goal} presents the experimental results on the Goal Planning task. 
Among the baselines, BERT and BERT+CNN perform much better than MGCG by fine-tuning a BERT-based model to encode the contextual information from the dialogue history, which attaches great importance in determining where the conversation should go at the next turn. 
Two variants of UniMIND achieve similar performance and both of them significantly outperform these baselines on two datasets.

\subsubsection{\textbf{Evaluation on Topic Prediction}}

\begin{table*}
    \caption{Evaluation of Topic Prediction on TG-ReDial. We report P@1, R@1, and F1@1 scores for the matching-based baselines, which achieve the highest F1@$k$ scores. We regard the first generated topic as the top-ranked topic to compute the Hit@1 score for UniMIND. $^\dagger$ indicates statistically significant improvement ($p$<0.05) over \underline{the best baseline}.}
    \centering
    \begin{tabular}{lcccccc}
    \toprule
        \multirow{2}{*}{Model} & \multicolumn{6}{c}{TG-ReDial}  \\
        \cmidrule(lr){2-7}
        & Hit@1 & Hit@3 & Hit@5 & P & R &F1 \\
        \midrule
        MGCG&0.3635& 0.5173& 0.6009& 0.5211&0.5013& 0.5079\\
        Topic-BERT&0.4381& 0.5823& 0.6246&0.6221& 0.5548& 0.5772\\
        Profile-BERT&0.0907& 0.1597& 0.2248& 0.3945&0.3740& 0.3808\\
        Conv-BERT&0.4348& 0.5873& 0.6329&0.6264& \underline{0.5589}& 0.5814\\
        Union&\underline{0.4420}&\underline{0.5923}&\underline{0.6374}&\underline{0.6301}& 0.5586& \underline{0.5824}\\
        \midrule
        \textbf{UniMIND}$_\text{N}$&0.7319$^\dagger$&-&-&0.6876$^\dagger$& 0.6915$^\dagger$& 0.6889$^\dagger$\\
        \textbf{UniMIND}$_\text{S}$&\textbf{0.7351}$^\dagger$&-&-&\textbf{0.6912}$^\dagger$& \textbf{0.6951}$^\dagger$& \textbf{0.6925}$^\dagger$\\
    \bottomrule
    \end{tabular}
    \label{exp:know_tg}
\end{table*}

\begin{table*}
    \caption{Evaluation of Topic Prediction on DuRecDial. We report P@1, R@1, and F1@1 scores for the matching-based baselines, which achieve the highest F1@$k$ scores. We regard the first generated topic as the top-ranked topic to compute the Hit@1 score for UniMIND. $^\dagger$ indicates statistically significant improvement ($p$<0.05) over \underline{the best baseline}.}
    \centering
    \begin{tabular}{lcccccc}
    \toprule
        \multirow{2}{*}{Model} & \multicolumn{6}{c}{DuRecDial} \\
        \cmidrule(lr){2-7}
        & Hit@1 & Hit@3 & Hit@5 & P & R &F1  \\
        \midrule
        MGCG&0.6639& 0.7510& 0.7810&0.6697&0.5419&  0.5880\\
        Topic-BERT&0.6337& 0.7460& 0.7785&0.6957& 0.6015& 0.6289\\
        Profile-BERT&0.4908& 0.6844& 0.7698&0.4576& 0.4036& 0.4181\\
        Conv-BERT&0.7791& 0.8226& 0.8425&0.8122& 0.7065& 0.7377\\
        Union&\underline{0.7877}& \underline{0.8462}& \underline{0.8696}&\underline{0.8327}& \underline{0.7270}& \underline{0.7582}\\
        \midrule
        \textbf{UniMIND}$_\text{N}$&\textbf{0.9056}$^\dagger$&-&-&\textbf{0.8981}$^\dagger$& \textbf{0.8994}$^\dagger$& \textbf{0.8978}$^\dagger$\\
        \textbf{UniMIND}$_\text{S}$&0.9023$^\dagger$&-&-&0.8957$^\dagger$& 0.8964$^\dagger$& 0.8952$^\dagger$\\
    \bottomrule
    \end{tabular}
    \label{exp:know_du}
\end{table*}

Table~\ref{exp:know_tg} and Table~\ref{exp:know_du} present the experimental results on the Topic Prediction task. 
Note that the results on TG-ReDial reported in \cite{tgredial} are based on the assumption that the target recommendation topic is given, so that all the baselines achieve a similar result, due to the strong supervision signal of the target topic. 
In fact, without the guidance of the target topic, the task of topic prediction becomes more difficult, since the system is required to predict the next topic based on the coherency and relevancy to the dialogue context, the historical topic thread, and the user profile. 
Therefore, we can observe that Profile-BERT barely works on the topic prediction task.  
Overall, UniMIND significantly outperforms these strong baselines on both Hit scores and F1 scores, where the Hit@1 scores of UniMIND are even higher than the Hit@5 scores of Union.  
Since it is difficult to determine the number of topic classes with these matching-based baselines, UniMIND owns remarkable flexibility and scalability to this multi-label classification problem, which is proven by the F1 scores.

\subsubsection{\textbf{Evaluation on Item Recommendation}}

\begin{table*}
    \caption{Evaluation of Item Recommendation on TG-ReDial. $^\dagger$ indicates statistically significant improvement ($p$<0.05) over \underline{the best baseline}.}
    \centering
    \begin{tabular}{lcccc}
    \toprule
        \multirow{2}{*}{Model} & \multicolumn{4}{c}{TG-ReDial}  \\
        \cmidrule(lr){2-5}
        & NDCG@10 & NDCG@50 & MRR@10 & MRR@50 \\
        \midrule
        GRU4Rec&0.0028&0.0062&0.0014&0.0020\\
        SASRec&0.0092&0.0179&0.0050&0.0068\\
        TextCNN&0.0144&0.0215&0.0119&0.0133\\
        BERT&0.0246&0.0439&0.0182&0.0211\\
        Union&\underline{0.0348}&\underline{0.0527}&\underline{0.0240}&\underline{0.0277}\\
        \midrule
        \textbf{UniMIND}$_\text{N}$&0.0306& 0.0499& 0.0236& 0.0277\\
        \textbf{UniMIND}$_\text{S}$&\textbf{0.0386}$^\dagger$& \textbf{0.0638}$^\dagger$& \textbf{0.0283}$^\dagger$& \textbf{0.0319}\\
    \bottomrule
    \end{tabular}
    \label{exp:rec_tg}
\end{table*}

\begin{table*}
    \caption{Evaluation of Item Recommendation on DuRecDial. $^\dagger$ indicates statistically significant improvement ($p$<0.05) over \underline{the best baseline}.}
    \centering
    \begin{tabular}{lcccc}
    \toprule
        \multirow{2}{*}{Model} & \multicolumn{4}{c}{DuRecDial} \\
        \cmidrule(lr){2-5}
        & NDCG@10 & NDCG@50 & MRR@10 & MRR@50 \\
        \midrule
        GRU4Rec&0.2188& 0.2734& 0.1713& 0.1833\\
        SASRec&0.3686& 0.4130& 0.3071& 0.3174\\
        TextCNN&0.5049& 0.5344& 0.4516& 0.4584\\
        BERT&0.5455& 0.5719& 0.4983& 0.5043\\
        Union&\underline{0.5568}& \underline{0.5831}& \underline{0.5101}& \underline{0.5159}\\
        \midrule
        \textbf{UniMIND}$_\text{N}$&0.5986$^\dagger$& 0.6099$^\dagger$& 0.5922$^\dagger$& 0.5944$^\dagger$\\
        \textbf{UniMIND}$_\text{S}$&\textbf{0.6343}$^\dagger$& \textbf{0.6471}$^\dagger$& \textbf{0.6291}$^\dagger$& \textbf{0.6318}$^\dagger$\\
    \bottomrule
    \end{tabular}
    \label{exp:rec_du}
\end{table*}

Table~\ref{exp:rec_tg} and Table~\ref{exp:rec_du} summarize the experimental results on the Item Recommendation task. 
Among these baselines, due to the sparsity of the historical user-item interaction data, traditional recommendation methods (\textit{e.g.}, GRU4Rec and SASRec) fall short of handling the item recommendation task in MG-CRS, while text-based recommendation methods (\textit{e.g.}, TextCNN and BERT) show more promising performance. Union further improves the performance by combining the advantages of BERT and SASRec. 
Without using the historical interaction data, UniMIND achieves competitive performance with Union. 
In specific, UniMIND$_\text{S}$ significantly and consistently outperforms Union on both datasets and UniMIND$_\text{N}$ outperforms Union on DuRecDial. 
Different from the observations on other tasks, UniMIND$_\text{S}$ performs much better than UniMIND$_\text{N}$ on item recommendation, showing that the natural language prompts can not fully utilize the relationships with the expanded item vocabulary. 

Overall, the experimental results provide the answer to \textbf{RQ2}: \textit{UniMIND significantly outperforms existing strong baselines on each sub-task of MG-CRS. The strong performance can not only contribute to the final quality of the generated responses, but also provide useful and convenient adaptation for different sub-task applications.}

\section{Detailed Analyses \& Discussions}
In this section, we provide a variety of detailed analyses and discussions to look deeper into the proposed method. Note that the following analyses are all conducted with UniMIND$_\text{S}$, since this variant has generally better performance on both datasets according to Section~\ref{sec:exp}.

\subsection{Ablation Study (RQ3)}
\begin{table}
    \caption{Ablation Study (Results on Response Generation).}
    \centering
    \begin{tabular}{lcccc}
    \toprule
        \multirow{2}{*}{Model} & \multicolumn{2}{c}{TG-ReDial} & \multicolumn{2}{c}{DuRecDial} \\
        \cmidrule(lr){2-3} \cmidrule(lr){4-5}
        & F1 & BLEU-1/2 & F1 & BLEU-1/2 \\
        \midrule
        \textbf{UniMIND}$_\text{S}$&35.62&0.314/0.090&51.87&0.477/0.397\\
        - w/o MTL&34.73&0.305/0.086&50.98&0.465/0.381\\
        - w/o PL &33.94&0.299/0.084&50.67&0.454/0.373\\%
        \midrule
        \multicolumn{5}{l}{\textit{Response Generation}}\\
        - OracleGen&39.43&0.350/0.104&55.07&0.525/0.444\\
        - DirectGen&32.80&0.291/0.070&48.41&0.418/0.328\\
        \midrule
        \multicolumn{5}{l}{\textit{Goal Planning}}\\
        - Oracle&36.78&0.328/0.097&53.28&0.498/0.417\\
        - BERT+CNN&35.51&0.314/0.090&51.67&0.475/0.396\\
        - w/o goal&35.13&0.309/0.086&50.97&0.467/0.380\\
        \midrule
        \multicolumn{5}{l}{\textit{Topic Prediction}}\\
        - Oracle&38.22&0.336/0.100&52.68&0.492/0.413\\
        - Union&34.32&0.301/0.084&51.09&0.469/0.383\\
        - w/o topic&35.01&0.307/0.085&48.73&0.426/0.340\\
        \midrule
        \multicolumn{5}{l}{\textit{Item Recommendation}}\\
        - Oracle&35.72&0.315/0.091&52.42&0.488/0.409\\
        - Union&35.60&0.314/0.090&51.25&0.471/0.394\\
        - w/o item&35.73&0.315/0.090&51.10&0.469/0.385\\
    \bottomrule
    \end{tabular}
    \label{exp:ablation}
\end{table}

In order to investigate the effect of the proposed training procedure and each task, the results of the ablation study are presented in Table~\ref{exp:ablation}. 

\subsubsection{Effect of Training Procedure}
We first evaluate the effectiveness of the multi-task learning and the prompt-based learning strategies. 
``- w/o MTL'' denotes that we train four independent Seq2Seq models for each task. 
``- w/o PL'' denotes that we only train one unified multi-task learning model for all tasks without task-specific prompt-based learning. 
\textit{The results show that the best performance can only be achieved by combining multi-task learning and prompt-based learning}, which can answer the first part of \textbf{RQ3}.

\subsubsection{Effect of Each Sub-task}
Besides, we also present the performance of ``- OracleGen'', which denotes that the input sequence for response generation is composed of the ground-truth goals, topics, and items, while ``- DirectGen'' denotes the input sequence only contains the dialogue history. 
The results show that the performance is improved by 15\%-25\% (DirectGen $\rightarrow$ OracleGen), which demonstrates the importance of these kinds of information in MG-CRS. 
In addition, we further investigate the effect of each task, by replacing them with the ground-truth labels (``- Oracle'') or predicted results from other strong baselines (``- BERT+CNN'' and ``- Union''), or discarding the information (``- w/o''). 
The results show that Topic Prediction and Goal Planning largely affect the final performance, where precisely predicting topics can bring the most prominent improvement on the final response generation. 
However, Item Recommendation has the least effect on the final response generation for both datasets.  

Overall, for the answer to the second part of \textbf{RQ3}, \textit{the experimental results show that all sub-tasks attach more or less importance to the final Response Generation task, among which Topic Prediction and Goal Planning are more influential than Item Recommendation.}

\subsection{Performance w.r.t. Goal Type (RQ4)}

\begin{table}
    \caption{Performance w.r.t. Goal Type.}
    \centering
    \setlength{\tabcolsep}{1.1mm}{
    \begin{tabular}{lcccccc}
    \toprule
    \multirow{2}{*}{Goal Type}&\multirow{2}{*}{\%}&Goal&Topic&\multicolumn{3}{c}{Response Gen.}\\
    \cmidrule(lr){3-3}\cmidrule(lr){4-4}\cmidrule(lr){5-7}
    &&F1&F1&F1&BLEU-1/2&Dist-2\\
    \midrule
    \multicolumn{7}{c}{TG-ReDial}\\
    \midrule
    Recommend.&54.4&\textbf{0.9629}&\textbf{0.8864}&37.6&0.337/0.072&0.218\\
    Chit-chat&39.0&0.9428&0.3886&30.5&0.254/0.071&\textbf{0.327}\\
    Rec. Request&31.9&0.8352&0.6926&\textbf{45.4}&\textbf{0.404/0.167}&0.251\\
    \midrule
    \multicolumn{7}{c}{DuRecDial}\\
    \midrule
    Recommend.&37.2&0.9235&0.7933&45.9&0.455/0.376&0.101\\
    Chit-chat&15.5&0.8734&0.9787&41.7&0.396/0.309&\textbf{0.132}\\
    QA&16.7&0.9298&0.9278&62.5&0.587/0.505&0.122\\
    Task&11.3&\textbf{0.9456}&\textbf{0.9963}&\textbf{68.5}&\textbf{0.701/0.637}&0.114\\
    \bottomrule
    \end{tabular}}
    \label{exp:goal_type}
\end{table}

In order to further analyze the characteristics of MG-CRS against traditional CRS, we present the performance with respect to different goal types in Table~\ref{exp:goal_type}. 
In order to provide a comprehensive scope, we aggregate the results of several dominant goal types for each dataset. 
For the TG-ReDial dataset that contains responses with multiple goals, the type-wise scores are averaged from all the samples that contain the goal type. 
For the DuRecDial dataset that contains multi-domain dialogues, the type-wise scores are averaged from all the samples that contain the goal type across all domains. 
On both datasets, ``\textit{Recommendation}'' is still the most important goal type with the largest number of samples in MG-CRS. 
There are several notable observations as follows: 
\begin{itemize}
    \item As for Goal Planning, ``\textit{Recommendation}'' is the easiest goal type to plan for TG-ReDial, since this type always comes after ``\textit{Recommendation Request}''. However, the timing for ``\textit{Recommendation Request}'' would be more difficult to determine. 
    \item As for Topic Prediction, the topic labels in TG-ReDial are the abstractive conversational topics, while those in DuRecDial are the knowledge entities discussed in the current conversation. 
    The difference in annotations leads to the different observations in the two datasets. 
    For ``\textit{Chit-chat}'' dialogues, the abstractive topics are hard to predict in TG-ReDial, while the discussed entities can be effectively predicted in DuRecDial. 
    Conversely, for ``\textit{Recommendation}'' dialogues, the abstractive topics are often the genre of the recommended movies in TG-ReDial, while there might be multiple knowledge entities related to the recommended item in DuRecDial. 
    \item As for Response Generation, ``\textit{Chit-chat}'' dialogues reach the lowest scores on the content preservation metrics (\textit{i.e.}, F1 and BLEU), while achieving the highest scores on the diversity metrics (\textit{i.e.}, Dist) in both datasets. 
    This phenomenon is prevalent in chit-chat dialogue studies. 
    ``\textit{Recommendation}'' dialogues reach the lowest scores on Dist, due to the similar expressions when making recommendations. 
\end{itemize}

Therefore, we can derive the answer to \textbf{RQ4} from this analysis: \textit{The performance w.r.t. different goal types demonstrate the difficulties of MG-CRS, since there are some great differences among different types of dialogues, not just making recommendations in traditional conversational recommender systems.}

\begin{figure}
\centering
\includegraphics[width=0.95\textwidth]{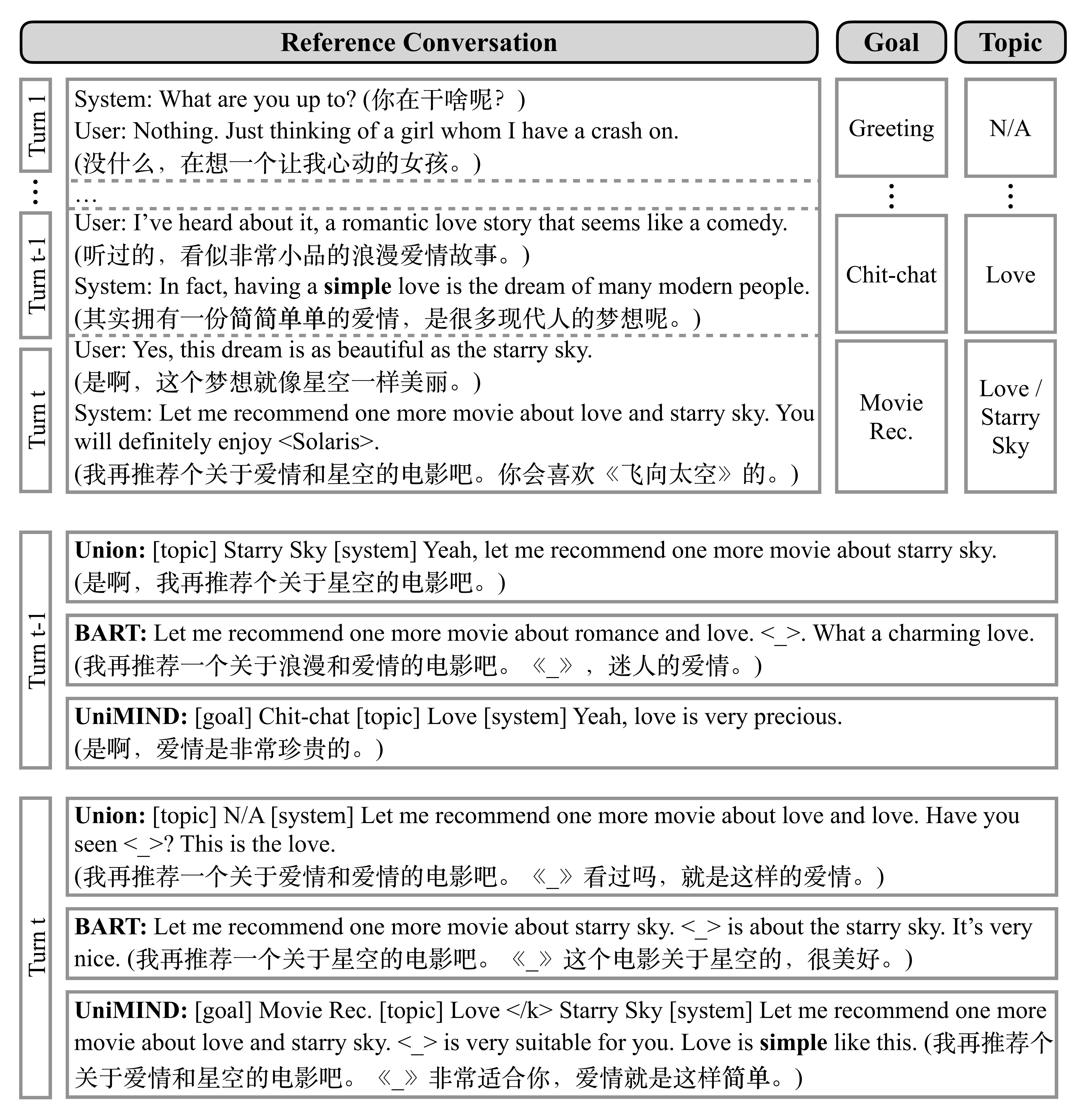}
\caption{Case Study.}
\label{case}
\end{figure}

\subsection{Case Study}

In order to intuitively differentiate UniMIND from other baselines, Fig.~\ref{case} presents a specific dialogue from TG-ReDial. 
At the $t$-$1$-th turn, the ground-truth conversational goal and topics are ``\textit{Chit-chat}'' and ``\textit{Love}'', which means that the system is expected to provide a causal response talking about love. 
However, Union predicts the topic to be ``\textit{Starry Sky}'' in advance, due to the strong supervision signals of the target topics of the movie to be recommended. 
Moreover, without the goal planning task, Union and BART tend to make recommendations frequently, since the majority of the conversational goals in MG-CRS is ``\textit{Recommendation}'' as shown in Table~\ref{exp:goal_type}. 
This is likely to degrade the user experience. 
UniMIND generates an appropriate response to discuss about love with the user, by making good goal planning and topic prediction. 
At the $t$-th turn, the ground-truth conversational goal and topics are ``\textit{Movie Recommendation}'' and ``\textit{Love/Starry Sky}'', which means that the system is expected to recommend a movie about love and starry sky. 
UniMIND can better capture the multiple topics and provide a more coherent response to the dialogue context.

\subsection{Error Analysis and Limitations}\label{sec:error}
Despite the effectiveness of the proposed UniMIND framework for MG-CRS, we would like to better understand the failure modes of UniMIND for further improvement in  future studies. 
After analyzing those cases with low human evaluation scores, we identify the following limitations and discuss the potential solutions:   
\begin{itemize}
    \item \textbf{Low Recommendation Success Rate}. All the baselines and UniMIND fail to reach a promising recommendation performance on TG-ReDial as shown in Table~\ref{exp:rec_tg}, due to the sparsity of the user-item interactions. Since the historical interaction data is not utilized in UniMIND, one possible direction is to study how to incorporate this kind of data into the Seq2Seq framework for improving the recommendation performance.  
    \item \textbf{Informativeness}. As shown in Table~\ref{exp:human}, there is still a gap between the generated and the ground-truth response on \textbf{Informativeness}. In order to diversify and enrich the information in dialogue systems, a common practice is to leverage open-domain dialogue corpus to post-train the generation model~\cite{dialogpt}, which can also be easily applied to our unified Seq2Seq framework. 
    \item \textbf{Error Propagation}. This is a typical issue of solving multiple tasks in sequential order. Table~\ref{error} presents the Exact Match (EM) scores between the generated input sequence and the oracle input sequence for each task, which inevitably go down along with the sequential completion of each task. There are some techniques studied to alleviate this issue in cascaded generation methods, such as introducing contrastive objectives~\cite{soloist} or noisy channel models~\cite{tacl21-noisy}.  
\end{itemize}

\begin{table}
    \caption{EM between generated and oracle input sequences.}
    \centering
    \begin{tabular}{lccc}
    \toprule
        Dataset & Topic Pred. & Item Rec. & Response Gen. \\
        \midrule
        TG-ReDial&87.20\%&80.17\%&15.44\%\\
        DuRecDial&92.32\%&89.16\%&84.74\%\\
    \bottomrule
    \end{tabular}
    \label{error}
\end{table}
\section{Conclusions}
In this work, we propose a novel unified multi-task learning framework for multi-goal conversational recommender systems, namely UniMIND. 
Specifically, we unify four tasks in MG-CRS into the same sequence-to-sequence (Seq2Seq) paradigm and utilize prompt-based learning strategies to endow the model with the capability of multi-task learning. 
Experimental results on two MG-CRS datasets show that the proposed method achieves state-of-the-art performance on each task with a unified model. 
Extensive analyses demonstrate the importance of each task and the difficulties of handling different types of dialogues in MG-CRS.

This work is the first attempt towards a unified multi-task learning framework for MG-CRS. There are some limitations and room for further improvement. 
As discussed in Section~\ref{sec:error}, the error analyses and limitation discussions shed some potential directions for future studies. For example, it can be beneficial to incorporate historical user-item interaction data into the unified framework for making better recommendations or leverage open-domain dialogue corpus to post-train the generation model for generating more informative and diverse responses. It would be also worth investigating approaches to alleviate the error propagation issue in the training procedure. 
In addition, similar to other prompt-based learning studies, the proposed method can be extended to handle the low-resource scenarios or few-shot learning settings in MG-CRS, which will be more practical in real-world applications.


\bibliographystyle{ACM-Reference-Format}

\bibliography{sample-base}

\end{document}